\documentclass[aps,prl,reprint,amssymb,showpacs,superscriptaddress,notitlepage,onecolumn]{revtex4-2}

\pdfoutput = 1
\usepackage{bm}
\usepackage{graphicx,hyperref}
\usepackage{dcolumn}
\usepackage{braket}
\usepackage{natbib}
\usepackage{amsmath}
\usepackage{xcolor}
\usepackage{mathtools,amssymb}
\usepackage{soul}
\usepackage{float}
\usepackage{amssymb}
\usepackage{esint}
\usepackage{mathtools}
\usepackage{physics}
\usepackage{makecell}
\usepackage{multirow}
\usepackage{array}
\usepackage{geometry}
\usepackage{url}
\usepackage{adjustbox}
\usepackage{rotating}
\usepackage{listings}
\usepackage{threeparttable,booktabs}
\usepackage{fontawesome}
\usepackage{cleveref}
\usepackage[version=4]{mhchem}
\renewcommand\hl[1]{#1} 

\begin{document}

\title{Observation of topological frequency combs}
\author{Christopher J. Flower}
\altaffiliation{These authors contributed equally to this work}
\affiliation{Joint Quantum Institute, National Institute of Standards and Technology/University of Maryland, College Park, MD, USA}
\author{Mahmoud Jalali Mehrabad}
\altaffiliation{These authors contributed equally to this work}
\affiliation{Joint Quantum Institute, National Institute of Standards and Technology/University of Maryland, College Park, MD, USA}
\author{Lida Xu}
\altaffiliation{These authors contributed equally to this work}
\affiliation{Joint Quantum Institute, National Institute of Standards and Technology/University of Maryland, College Park, MD, USA}
\author{Gregory Moille}
\affiliation{Joint Quantum Institute, National Institute of Standards and Technology/University of Maryland, College Park, MD, USA}
\author{Daniel G. Suarez-Forero}
\affiliation{Joint Quantum Institute, National Institute of Standards and Technology/University of Maryland, College Park, MD, USA}
\author{Oğulcan Örsel} 
\affiliation{Department of Mechanical Science and Engineering, University of Illinois at Urbana-Champaign, Urbana, IL, USA}
\author{Gaurav Bahl} 
\affiliation{Department of Mechanical Science and Engineering, University of Illinois at Urbana-Champaign, Urbana, IL, USA}
\author{Yanne Chembo} 
\affiliation{Institute for Research in Electronics and Applied Physics, University of Maryland, College Park, MD, USA}
\author{Kartik Srinivasan}
\affiliation{Joint Quantum Institute, National Institute of Standards and Technology/University of Maryland, College Park, MD, USA}
\author{Sunil Mittal}\email{s.mittal@northeastern.edu}
\affiliation{Department of Electrical and Computer Engineering, Northeastern University, Boston, MA, USA}
\author{Mohammad Hafezi}
\email{hafezi@umd.edu}
\affiliation{Joint Quantum Institute, National Institute of Standards and Technology/University of Maryland, College Park, MD, USA}


\begin{abstract}
On-chip generation of optical frequency combs using nonlinear ring resonators has opened the route to numerous novel applications of combs that were otherwise limited to mode-locked laser systems. Nevertheless, even after more than a decade of development, on-chip nonlinear combs still predominantly rely on the use of single-ring resonators. Recent theoretical investigations have shown that generating combs in a topological array of resonators can provide a new avenue to engineer comb spectra. Here, we experimentally demonstrate the generation of such a novel class of frequency combs, topological frequency combs, in a two-dimensional (2D) lattice of hundreds of nonlinear ring resonators. Specifically, the lattice hosts topological edge states that exhibit fabrication-robust linear dispersion and spatial confinement at the boundary of the lattice. Upon optical pumping of the topological edge band, these unique properties of the edge states lead to the generation of a \hl{nested} frequency comb that is spectrally confined within the edge bands across $\approx$40 longitudinal modes. Moreover, using spatial imaging of our topological lattice, we confirm that light generated in the comb teeth is indeed spatially confined at the lattice edge, characteristic of linear topological systems. Our results bring together the fields of topological photonics and optical frequency combs, providing an opportunity to explore the interplay between topology and nonlinear systems in a platform compatible with commercially available nanofabrication processes.
\end{abstract}

\maketitle

\section{Introduction}

Nonlinear effects, in particular the Kerr effect, in ring resonators provide a compact route to the generation of optical frequency combs in integrated photonic chips \cite{Kippenberg2011, Pasquazi2018, Gaeta2019,diddams2020optical}. These combs have led to a plethora of applications including spectroscopy \cite{Suh2016, long2023nanosecond}, precision timekeeping, on-chip signal synthesis, ranging and detection \cite{Riemensberger2020, chen2023breaking}, and optical neural networks. While Kerr combs have been demonstrated in a wide variety of integrated material platforms, device design has been predominantly limited to single-ring resonators. Coupled resonator systems have only very recently been investigated as a means to engineer the dispersion, and subsequently, the comb spectrum \cite{Miller2015,Kim2017DispEng, Bao2019, Xue2019, yuan2023soliton}. Beyond dispersion engineering, nonlinear coupled resonator systems can exhibit  coherent solutions that are not possible using single-ring resonators \cite{Vasco2019, tikan2021emergent, Tusnin2023,yuan2023soliton}.  

Concurrently, topological photonics has emerged as a new and powerful paradigm for the design of photonic devices with novel functionalities \cite{LingLu2014, ozawa2019topological, price2022roadmap,zhang2023second,smirnova2020nonlinear,jalali2023topological}. More specifically, topological systems exhibit chiral or helical edge states that are confined to the boundary of the system and are remarkably robust against imperfections common to integrated photonic devices \cite{Wang2009, Rechtsman2013,Hafezi2013}. Examples include robust optical delay lines \cite{mittal2014topologically}, chiral quantum optics interfaces \cite{barik2018topological,mehrabad2023chiral,guddala2021topological}, slow light engineering \cite{guglielmon2019broadband}, waveguides, tapers and reconfigurable routers \cite{shalaev2019robust,Flower2023,zhao2019non}. While early efforts in topological photonics focused on linear devices, more recent demonstrations have included nonlinear effects, extending the scope of possible applications to include lasers \cite{St-Jean2017, Bahari2017, Bandres2018, Yang2022}, parametric amplifiers \cite{peano2016topological,sohn2022topological} and quantum light sources \cite{mittal2018topological, Blanco-Redondo2018, Mittal2021, Dai2022}. Additionally, it was theoretically shown that nonlinear effects in large two-dimensional (2D) topological arrays of ring resonators can lead to the generation of coherent nested temporal solitons that can exhibit an order of magnitude higher efficiency compared to single-ring combs \cite{mittal2021topological}.

 Here we experimentally demonstrate the generation of the first topological frequency comb in a 2D lattice of more than $100$ ring resonators, fabricated using a commercially available integrated Silicon Nitride (SiN) nanophotonic platform. As we pump within a topological edge band, we observe the generation of a frequency comb confined within the edge bands across $\approx$40 longitudinal modes. \hl{Using an ultra-high resolution spectrum analyzer, we reveal the unique nested structure of the comb where each comb tooth is further split into a set of finer teeth.} 
Furthermore, we directly image a set of comb teeth and verify that their spatial profile is indeed confined to the edge of the lattice. As such, they constitute a topological edge state that is robust against $90^\circ$ bends in the lattice and demonstrate the preservation of topology in a highly nonlinear system. This novel modulation instability comb is the first example of a new family of frequency combs and paves the way for the development of coherent topological frequency combs and nested temporal solitons \cite{mittal2021topological}.


\section{Design}

\begin{figure}[t]
  \centering
  \makebox[0pt]{\includegraphics[width=0.99\textwidth]{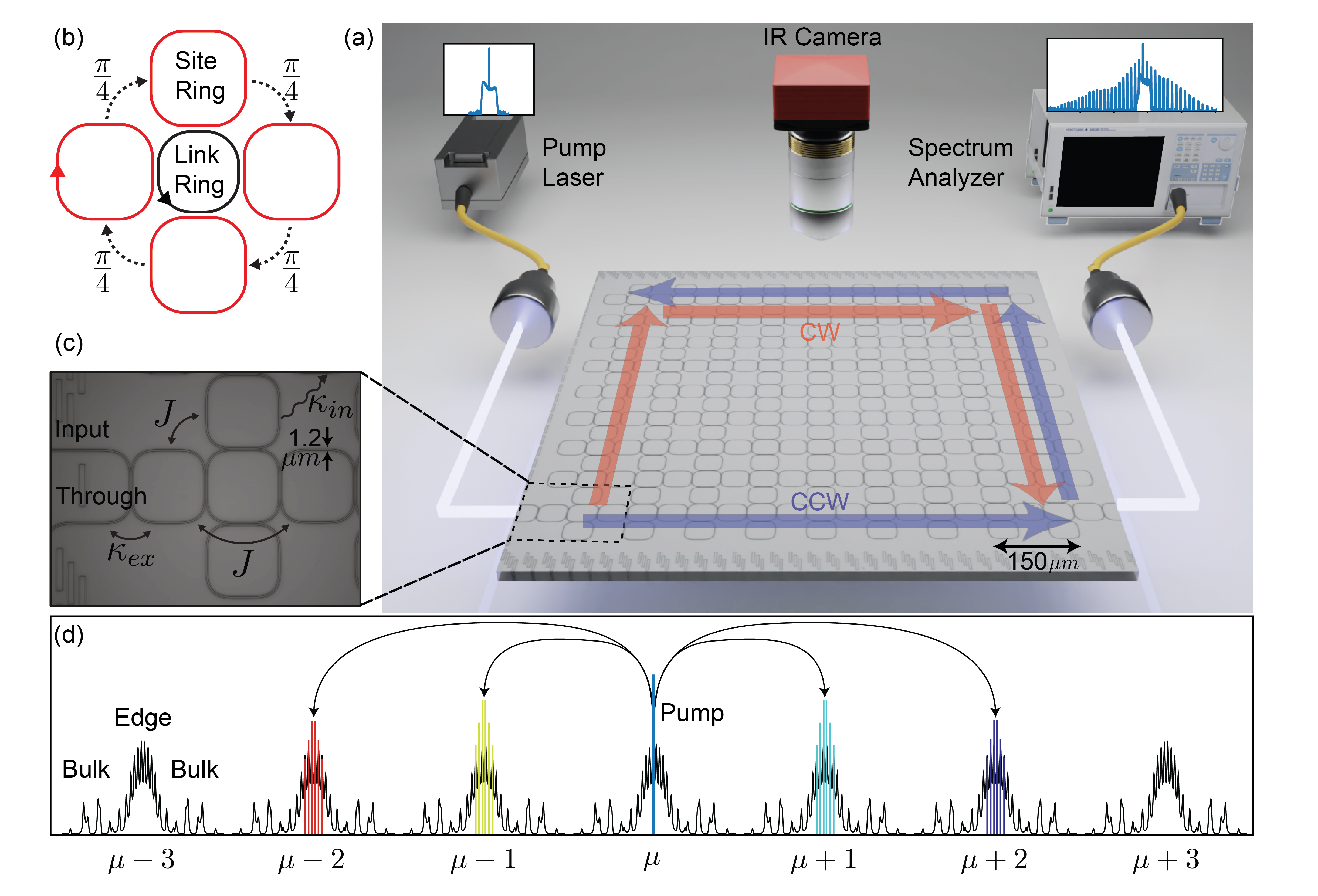}}
  \caption{\textbf{Generation of the topological frequency comb.} (a) Schematic of the pump, the spectral measurement setup, and an optical image of the device. A tunable pump is coupled into the lattice at the input port and circulates around the edge of the 2D AQH SiN lattice. The generated topological frequency comb spectra are collected from the drop port and analyzed with a spectrum analyzer. The paths followed by CW and CCW edge modes are highlighted in red and blue, respectively. (b) Photons acquire a non-zero phase ${\pi}/{4}$ when they hop to an adjacent site-ring (red) via a link-ring (black). (c) Close-up high-resolution optical image of the fabricated AQH lattice. Parameters and the input and through ports are marked. (d) Simulated linear transmission of the device with the four-wave mixing process schematically depicted.}
  \label{fig:schematic}
\end{figure}

Our topological system consists of an array of coupled ring resonators that simulates the anomalous quantum Hall (AQH) model for photons \cite{leykam2018reconfigurable,mittal2019photonic,mittal2021topological}, as shown in Figure~\ref{fig:schematic}. The \hl{180} ``site-ring" resonators form a square lattice, where nearest and next-nearest sites are coupled together via an interspersed lattice of \hl{81} detuned ``link-ring'' resonators. The link-rings are detuned by engineering a path-length difference with respect to the site-rings, ensuring that close to site-ring resonances the intensity present in the link-rings will be negligible. As a result, the link-rings act as waveguides and introduce a direction-dependent hopping phase of $\pm \pi/4$ for nearest-neighbor couplings and $0$ for next-nearest neighbor couplings. We note that our system implements a copy of the AQH model at each of the longitudinal mode resonances $\omega_{0,\mu}$ of the ring resonators. Therefore, the linear dynamics of the system are described by a multi-band tight-binding Hamiltonian ($\hat{H}_L$):

\begin{equation}
    \hat{H}_{ \rm L}=\sum_{m, \mu} \omega_{0,\mu} \hat{a}_{m, \mu}^{\dagger} \hat{a}_{m, \mu}-J \sum_{\langle m, n\rangle, \mu} \hat{a}_{m, \mu}^{\dagger} \hat{a}_{n, \mu} e^{-i \phi_{m, n}}-J \sum_{\langle\langle m, n\rangle\rangle, \mu} \hat{a}_{m, \mu}^{\dagger} \hat{a}_{n, \mu}.
    \label{eqn:linear Hamiltonian}
\end{equation}

Here $\hat{a}_{m, \mu}^{\dagger}$ is the photon creation operator at a site-ring $m$ and longitudinal mode $\mu$. The coupling strength between site-rings for both nearest and next-nearest neighbors is given by $J$. The resonance frequency of the site-ring resonators for a longitudinal mode with index $\mu$ is denoted: $\omega_{0,\mu}=\omega_0+D_1 \mu+\frac{D_2}{2}\mu^2, $
 where $D_1$ is the free spectral range (FSR), $D_2$ is the second-order dispersion\hl{, and $\omega_0$ is the resonance frequency of the pumped longitudinal mode, $\mu =0$.}
 This Hamiltonian leads to the existence of an edge band, spectrally located between two bulk bands, near each of the longitudinal mode resonances $\omega_{0,\mu}$, as shown in Figure~\ref{fig:schematic}. Furthermore, simulated transmission shows each edge band hosting another set of resonances. These resonances are associated with the longitudinal modes of the super-ring resonator formed by the edge states, giving rise to a nested mode structure.

We also emphasize that the system is time-reversal invariant and the topological edge states are helical in nature. More specifically, the clockwise (CW) and counterclockwise (CCW) circulation of light in the site-rings (also referred to as the pseudospin) leads to edge states that are circulating around the lattice boundary in the CCW and CW directions, respectively. By choosing the port of excitation, we can selectively excite a given edge state (Figure~\ref{fig:schematic}).

In the presence of a strong pump, the intrinsic Kerr nonlinearity of SiN leads to four-wave mixing, and subsequently, the generation of optical frequency combs in the lattice. This nonlinear interaction is described by the following Hamiltonian:

\begin{equation}
    \hat{H}_{\rm NL} = - \beta\sum_{m,\mu} \hat{a}^\dagger_{m,\mu_1} \hat{a}^\dagger_{m,\mu_2} \hat{a}_{m,\mu3} \hat{a}_{m,\mu4} \delta_{\mu_1+\mu_2,\mu_3+\mu_4},
    \label{eqn:nonlinear Hamiltonian}
\end{equation}

\noindent
where $\beta$ is the interaction strength between photons (see SI for definition). Finally, we note that our lattice is a driven-dissipative system. The dissipation includes both the intrinsic decay rate $\kappa_{\rm in}$ in each individual site, and the extrinsic decay rate $\kappa_{\rm ex}$ introduced by the coupling between input/output rings and probe waveguides, as shown Figure~\ref{fig:schematic}. The nonlinear dynamics of this coupled resonator system is described by a modified Lugiato-Lefever formalism which predicts the formation of nested optical frequency combs \cite{mittal2021topological}. In particular, pumping the lattice in one of the edge bands leads to efficient light generation only in the edge bands centered around other resonances $\omega_{0,\mu}$. This is because the spatial confinement of the edge modes ensures significant spatial overlap between different edge modes while minimizing overlap between edge and bulk modes.  Furthermore, the linear dispersion of the edge modes ensures that the anomalous dispersion from the waveguides is the dominant contribution as is typical with single-ring combs \cite{mittal2021topological,mittal2018topological}. 

\section{Device Fabrication}
This model is experimentally realized in a thick silicon nitride platform patterned through deep-UV lithography in a commercial foundry~\cite{thick_SiN}. The device itself consists of a 2D array of 261 coupled photonic ring resonators with two coupled input-output waveguides. A high-resolution optical image in Figure~\ref{fig:schematic} shows the topological photonic lattice used in this work. The waveguides are embedded in silicon dioxide and its dimensions are chosen to be 1200 nm wide by 800 nm thick in order to operate in the anomalous dispersion regime. Simulated mode profiles and dispersion can be found in the SI \cite{LukeSiN}. Each ring is a racetrack design, composed of $12$ $\mu$m straight coupling regions and $90^{\circ}$ Euler bend regions with a $20$ $\mu$m effective radius, giving rise to an FSR of $\approx0.75$ THz. Here, we specifically use Euler bends as opposed to round bends to reduce mode mixing that occurs at the straight-bent interfaces within each ring \cite{fujisawa2017low}. While this mode mixing and its impact in perturbing dispersion has also been of concern for single-racetrack combs \cite{ji2022Euler}, in our coupled resonator lattice its impact can be physically distinct. In particular, spurious hopping phases can be generated via mode conversion during hopping between adjacent rings \cite{tzuang2014non}. \hl{We also note that previous implementations of such topological devices have avoided this problem by operating in the single-mode regime} \cite{mittal2019photonic}. \hl{The constraints placed on waveguide geometry to access anomalous dispersion necessitate the use of wider waveguides that support higher-order modes}.

The coupling gaps between the resonators, as well as those between the input-output waveguides and the resonators, are 300 nm, corresponding to an approximate value of $2\pi\times25$ GHz for the coupling strength, $J$. The extrinsic and intrinsic couplings ($\kappa_{\rm ex}$ and $\kappa_{\rm in}$) are estimated to be $2\pi\times30$ GHz and $2\pi\times2$ GHz, respectively. For details on these calculations, see the SI. 


\section{Linear Measurements}
\begin{figure}[t]
  \centering
  \includegraphics[width=0.99\textwidth]{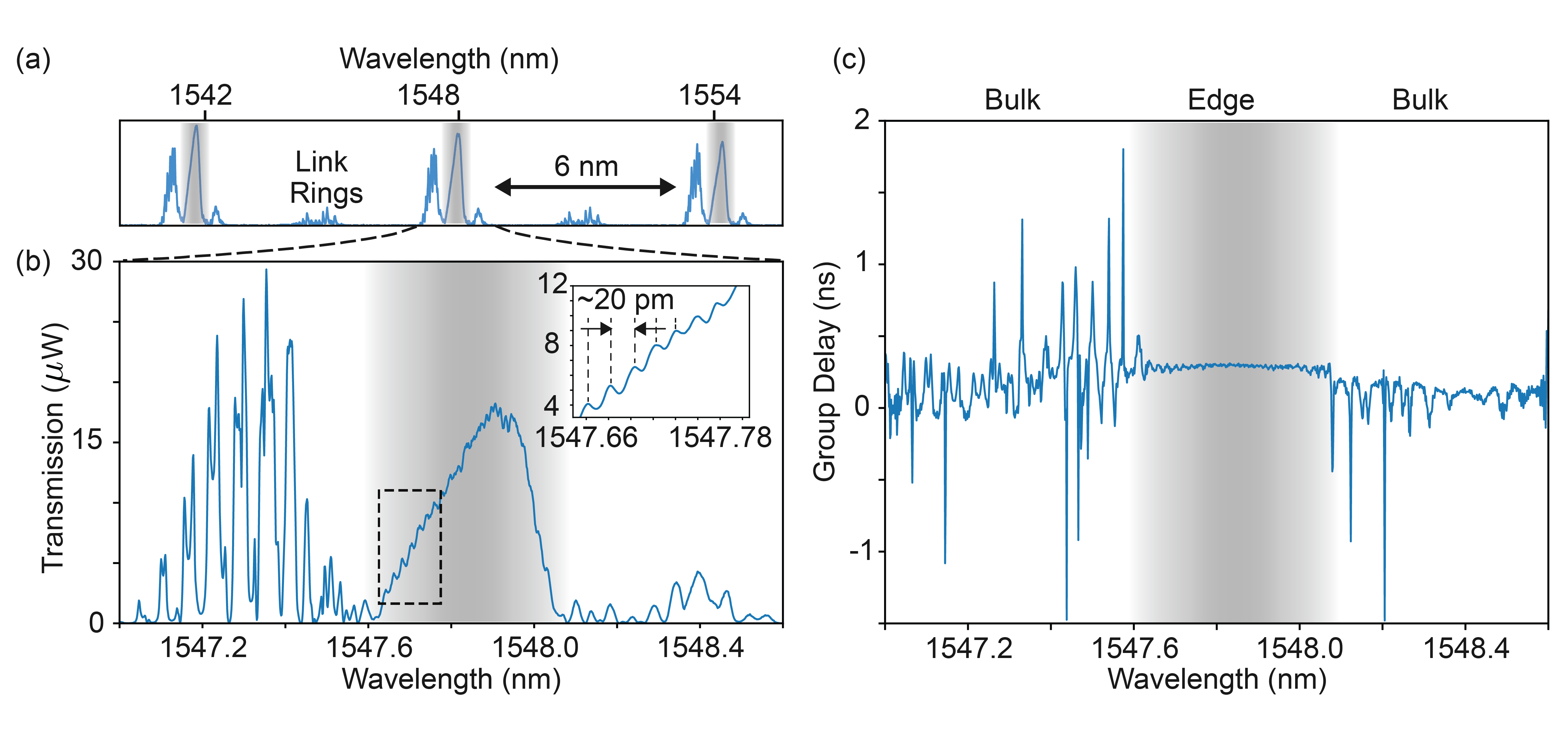}
  \caption{\textbf{Experimental characterization of the topological lattice.} (a) Measured drop transmission spectrum of the topological lattice showing bulk and edge bands for three longitudinal modes and detuned link-ring resonances. (b) Zoomed drop spectrum of the topological lattice on one set of edge and bulk bands. (c) The group delay spectrum showing a flat edge band.}
  \label{fig:linear}
\end{figure}

We begin by characterizing the transmission spectrum of the device in the linear (low-power) regime where Kerr phase shifts are negligible. Figure \ref{fig:linear} shows the measured \hl{drop port} transmission spectrum of the device over three longitudinal modes of the site-rings, \hl{as well as a single higher resolution spectrum across one transmission band. The edge bands are shaded in grey, and the spacing of the individual edge modes is highlighted in a zoomed inset.} While the topological edge states are robust against disorders, the bulk states are prone to reduced transmission. We also note that while the individual edge state resonances in our experiment are not well resolved due to the fact that the lattice is strongly coupled to input-output waveguides\hl{, the edge mode splitting can be approximated as 20 pm. This is in agreement with the estimated edge bandwidth of the device divided by the number of individual edge modes.}

To highlight the linear dispersion of edge states, we measure the group delay through the lattice, displayed in Figure~\ref{fig:linear},  \hl{using an optical vector analyzer}. As expected, the linear dispersion of the edge states leads to a flat \hl{group} delay response in the edge band, \hl{indicating  that the dispersion of the super-modes is small, and therefore, the device dispersion will be dominated by the single-ring dispersion.} The group delay through the bulk states, which do not have a well-defined momentum, shows prominent variations throughout the bulk band. 

\begin{figure}[t]
  \centering
  \makebox[0pt]{\includegraphics[width=0.99\textwidth]{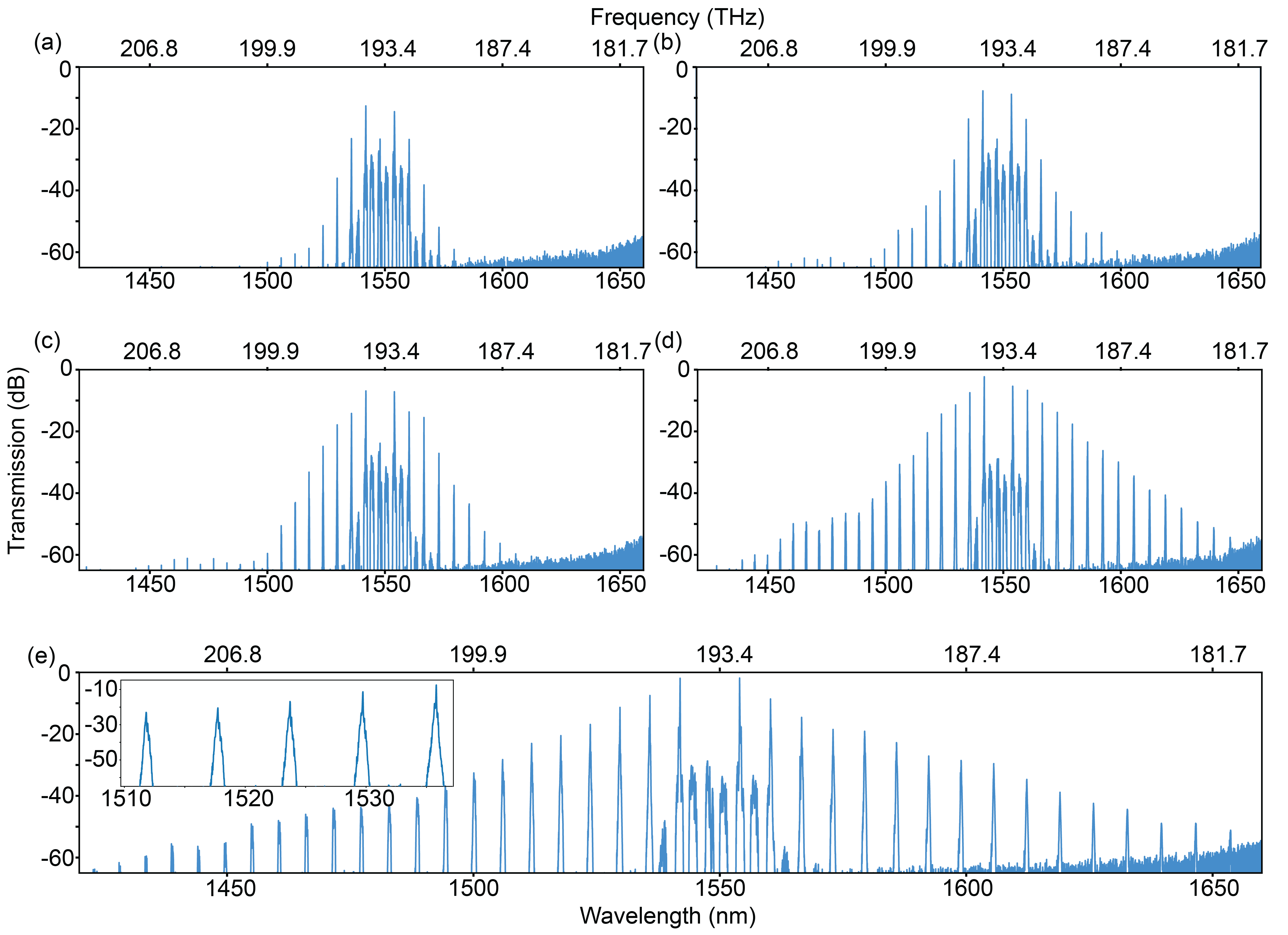}}
  \caption{\textbf{Formation of the topological frequency comb.} (a-e) Comb spectra measured with a pump laser wavelength of 1547.97 nm and peak powers of approximately \{70, 78, 86, 92, 100\} W, respectively. The inset on the left of (e) shows a zoomed spectrum of five comb teeth. }
  \label{fig:nonlinear}
\end{figure}

\begin{figure}[t]
  \centering
  \makebox[0pt]{\includegraphics[width=0.99\textwidth]{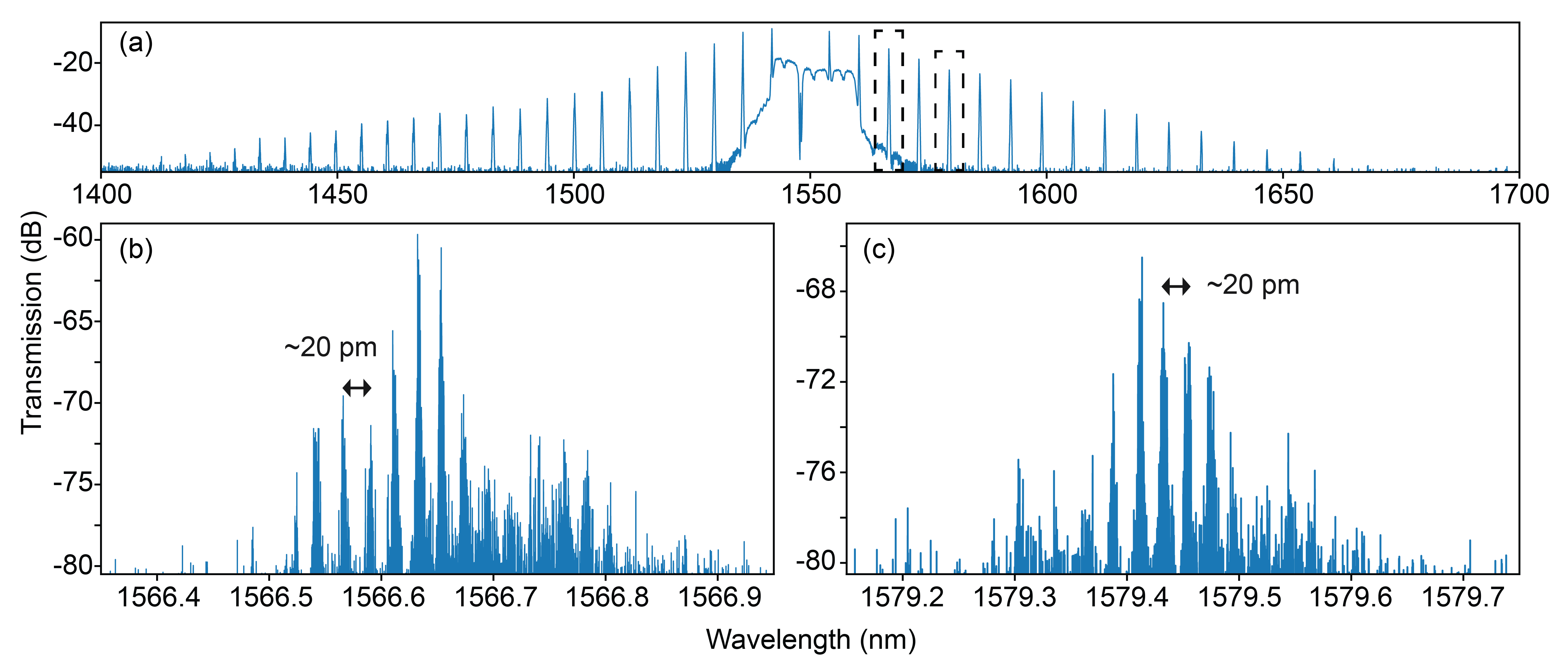}}
  \caption{\hl{\textbf{High-Resolution Spectra of Individual Comb Teeth.} (a) Low-resolution, broadband comb spectrum with a pump laser wavelength of 1547.97 and on-chip peak power of approximately 85 W. Individual comb teeth selected for high-resolution analysis are indicated with dotted lines. (b-c) High-resolution spectra of individual comb teeth showing a nested substructure around 1566.6 and 1579.4 nm respectively.}}
  \label{fig:nesting}
\end{figure}

\section{Nonlinear Measurements} 
To observe the formation of topological frequency combs in the ring resonator array, we pump the array using a 5 nanosecond pulsed laser with a repetition rate of 250 kHz and on-chip peak powers up to $\approx$100 W. We specifically choose a long pulse laser with a low duty cycle such that we can achieve a high peak power while keeping the average power low enough to avoid serious thermal effects. Furthermore, the 5 ns pulse duration is longer than any relevant timescale of the lattice dynamics, including the round-trip time in the super-ring resonator ($\approx$~\hl{400} ps). In other words, the longer pulse duration facilitates a selective quasi-continuous wave excitation of the edge band. \hl{We note that operating in this regime is necessary due to the unique challenges of fabricating a resonant structure of this scale. See SI for additional details of these challenges, the nonlinear measurement setup, and the pump laser spectrum.}

We pump the system at the edge band and show the emergence of the topological frequency comb as a function of increasing pump power. In particular, the \hl{drop port} spectra displayed in Figure~\ref{fig:nonlinear} were taken with a pump wavelength of 1547.97 nm and on-chip peak pump powers of 70, 78, 86, 92, 100 W. \hl{We estimate the threshold peak pump power to be $\approx$ 70 W, but also note that the threshold power changes with the pump wavelength.} The full comb bandwidth at the highest pump power is approximately 250 nm wide with about 65 dB contrast from the most prominent sidebands to the noise floor of the measurement. The inset of panel (e) shows a zoomed region of the spectrum around 1524 nm, spanning five longitudinal modes. The observed FSR of the comb is approximately 6 nm, in agreement with the single-ring FSR. \hl{For the broadest comb, the contrast between the height of the pump laser and the most prominent sideband (shown in the SI) is approximately 2.9 dB.}

\hl{To show the nested structure of the topological frequency comb within each comb tooth, we measure the comb output at the through port using an ultra-high resolution (0.04 pm) heterodyne-based optical spectrum analyzer. A reference low-resolution spectrum is shown in Figure}~\ref{fig:nesting}a. \hl{We select two individual comb teeth, as indicated, for high-resolution analysis. Within each of these comb teeth, we observe the oscillation of another set of well-resolved modes that correspond to the individual edge modes, shown in Figure}~\ref{fig:nesting}(b-c).\hl{ The spacing between the oscillating edge modes is about 20 pm, which corresponds to the free spectral range of the super-ring formed by the edge states and agrees with linear measurements. The linewidths of the individual edge modes vary in the range 3-5 pm. For comparison to bulk and single-racetrack comb spectra that lack this nested structure, see the SI.}

\begin{figure}[t]
  \centering
  \makebox[0pt]{\includegraphics[width=0.99\textwidth]{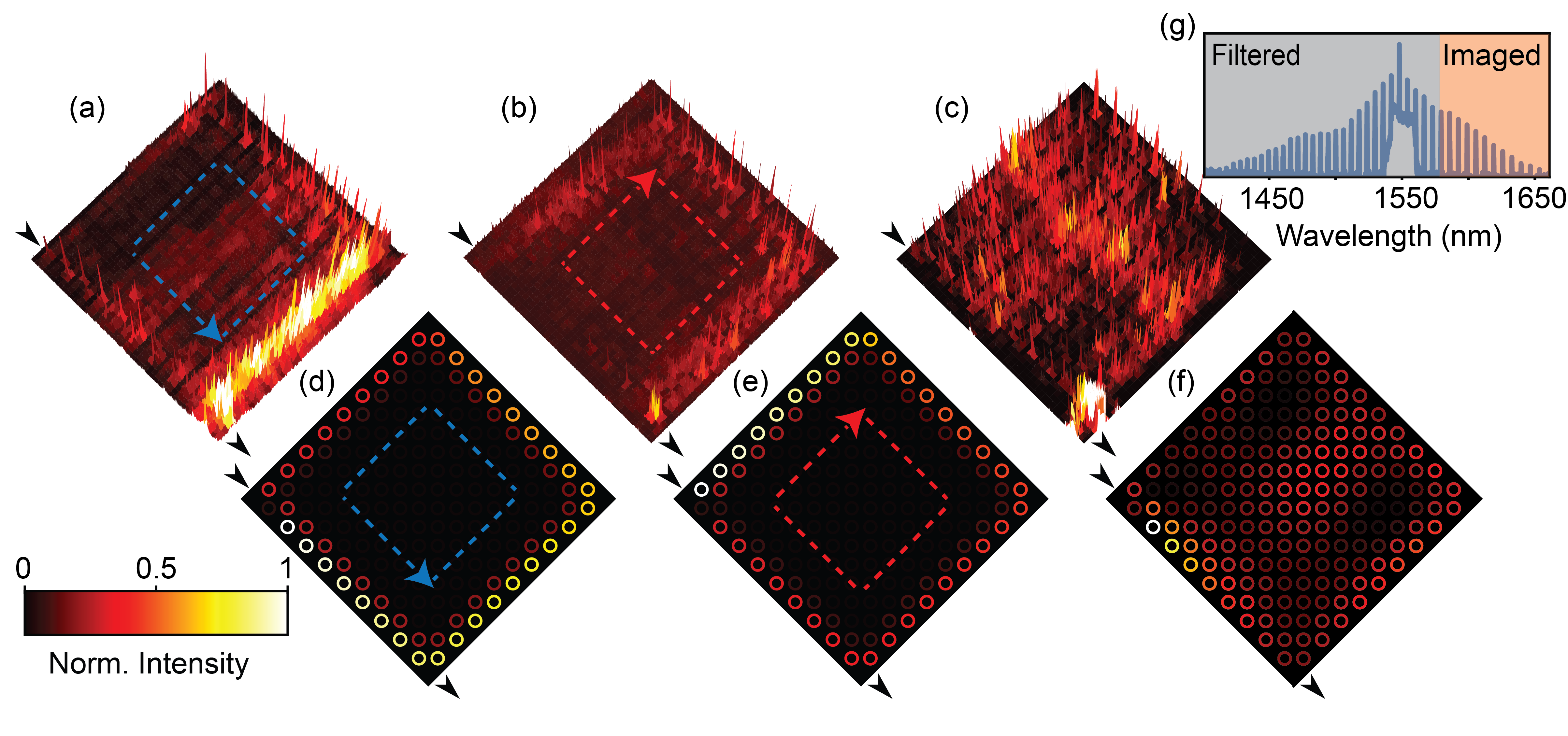}}
  \caption{\textbf{Spatial imaging of the topological frequency comb.} (a-c) Measured spatial imaging of the topological frequency comb for CCW, CW, and bulk modes, respectively. (d-f) Simulated spatial profile of the CCW, CW, and bulk modes in the linear regime.  (g) The integration bandwidth used for the top-down imaging.}
  \label{fig:imaging}
\end{figure}

To show that the topological frequency comb inherits the topological properties of the linear system and is indeed confined to the boundary of the lattice, we perform direct imaging of the generated comb.  While the system is designed to be well confined in plane, there is a certain amount of out-of-plane scattering caused by fabrication imperfections and disorder. The light scattered due to surface roughness is collected from above with a 10x objective lens and imaged on an infrared (IR) camera. In addition, we use a 1580 nm long-pass filter to remove the pump and only collect part of the generated comb light. 

Figure~\ref{fig:imaging} shows the measured spatial intensity profile of three types of generated frequency combs. First, we observe that the generated comb light is confined to the edge of the lattice and light travels from the input to the output port in the CCW direction. Note that the lattice is not critically coupled to the bus waveguides, therefore the light continues to circulate around the lattice after reaching the first output port in its path. Moreover, the propagation is robust and no noticeable scattering into the bulk is observed from the two sharp 90$^\circ$ corners. These characteristics show that the comb teeth are indeed generated within the topological edge band and that the topology is preserved even in the presence of strong nonlinearity.

Next, by pumping the system in the other pseudospin, we generate the comb in the CW edge state. We observe similar confinement of the topological frequency comb, but here the light travels in the opposite direction around the lattice, as expected. 

In sharp contrast to the CW and CCW edge band excitation, when we excite the lattice in the bulk band, the spatial intensity distribution of generated frequencies exhibits no confinement and occupies the bulk of the lattice. \hl{These images represent a novel look into the spatial profile of frequency comb formation, enabled by the unique geometry and scale of the topological lattice.} For details on the generation of each of these types of frequency combs, see the SI. Additionally, Figure~\ref{fig:imaging} shows simulated spatial distributions of CCW, CW, and bulk modes in the linear regime for comparison, as well as a schematic illustrating the filtered and imaged regions of the spectrum. \hl{We note that in these linear simulation results, we observe a uniform decay in intensity due to propagation loss from input to the output port. In contrast, our experimentally observed intensity profiles do not show a uniform decay, likely due to competition between the linear loss and the nonlinear parametric gain.}


\section{Outlook}
 Here we have demonstrated the first topological frequency comb using an array of more than $100$ coupled resonators. Our results entail the first realization of a new class of frequency combs that also includes coherent dissipative solutions, such as nested solitons and phase-locked Turing rolls, that are not accessible using single resonators \cite{mittal2021topological}. \hl{The unique nested spectral structure of these combs, characterized by two disparate frequency scales, could lead to a host of new applications. For instance,  this could be useful in certain spectroscopic measurements where there are multiple regions of interest that each require a high-resolution analysis but are separated by a large frequency gap. Moreover,} in this work we have used a commercially available silicon nitride platform in order to operate in the telecom wavelength regime. However, our device design can be easily translated to other frequency domains and photonic material platforms that can exhibit much higher nonlinearities such as aluminum gallium arsenide \cite{chang2020ultra,pu2016efficient} and lithium niobate \cite{zhang2019broadband}.

 On a more fundamental level, our results provide a new platform to study the interplay of topology and optical nonlinearities \cite{smirnova2020nonlinear, jurgensen2021quantized,mostaan2022quantized}, as well as intriguing topological physics unique to bosons. For example, while optical nonlinearities have been used to demonstrate topological phase transitions and restructured bulk-edge correspondence \cite{Maczewsky2020, Kirsch2021}, here we observe that the system retains the topological behavior of its linear counterpart even in the presence of such strong nonlinear effects. These results could enable novel applications where topological physics is used to engineer the underlying bandstructure (or dispersion) of a linear system and optical nonlinearities provide additional functionalities.



\bibliography{main.bib}


\section{Acknowledgements}
The authors wish to acknowledge fruitful discussions with Elizabeth Goldschmidt, Avik Dutt, Rahul Vasanth, Deric Session and Erik Mechtel. \textbf{Funding:} This work was supported by Airforce Office of Scientific Research FA9550-22-1-0339, Office of Naval Research N00014-20-1-2325, Army Research Lab W911NF1920181, National Science Foundation DMR-2019444, and Minta Martin and Simons Foundations. \textbf{Authors contribution:} C.J.F. designed the devices and constructed the experimental setup. C.J.F. performed the measurements with assistance from L.X. and M.J.M. C.J.F. and L.X. performed the simulations. C.J.F., L.X., and M.J.M analyzed the data. All work was supervised by G.M., K.S., S.M., and M.H. All authors discussed the results and contributed to the manuscript.  \textbf{Competing interests:} The authors declare no competing interests. \textbf{Data and materials availability:} All data needed to evaluate the conclusions in the
paper are present in the figures of the paper and/or the
supplementary materials. The data can be accessed at Dryad \cite{data}.



\newpage
\setcounter{equation}{0}
\setcounter{figure}{0}
\setcounter{table}{0}
\setcounter{page}{1}
\makeatletter
\renewcommand{\theequation}{S\arabic{equation}}
\renewcommand{\thefigure}{S\arabic{figure}}
\pagenumbering{roman}
\section{Supplementary Information}

\subsection{Band Structure}
\begin{figure}[h]
  \centering
  \includegraphics[width=0.99\textwidth]{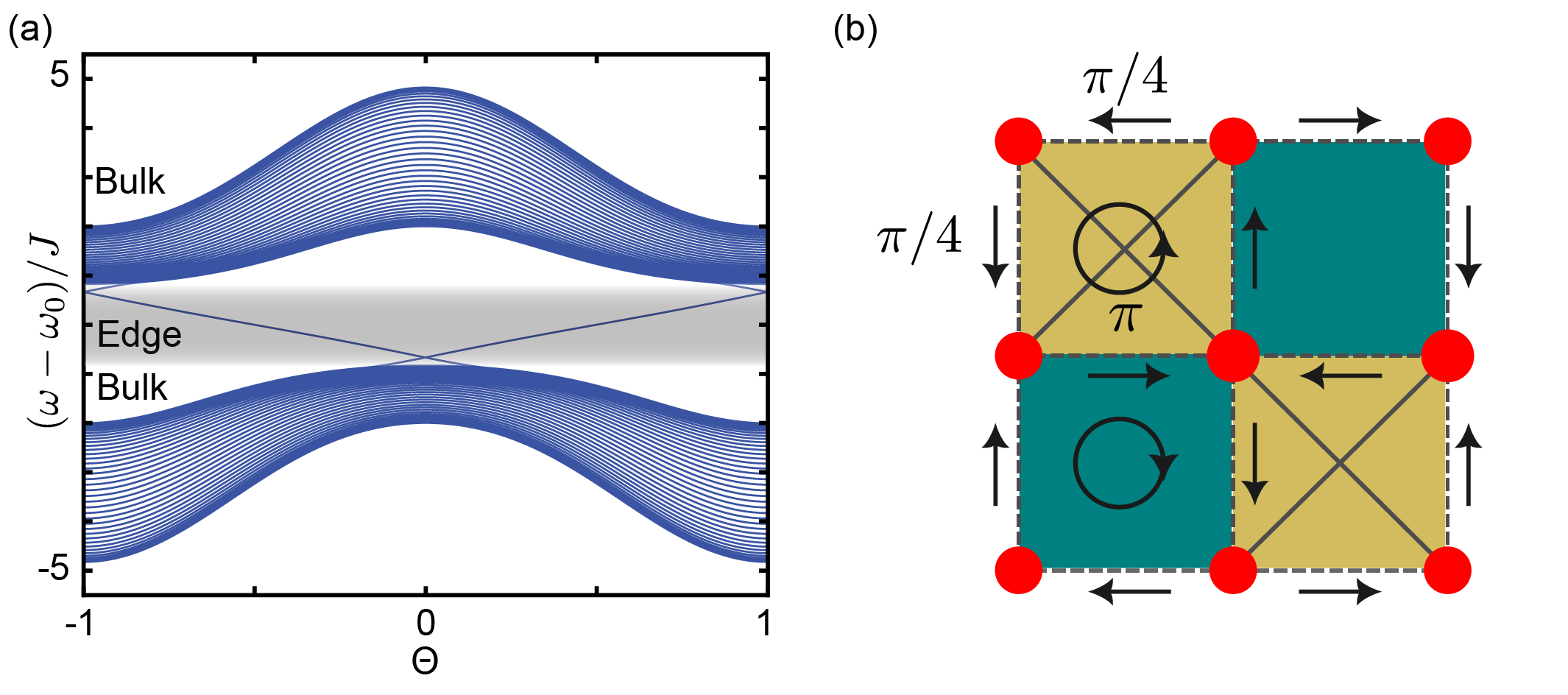}
  \caption{\textbf{Bandstructure.} (a) The band structure of a semi-infinite lattice (finite along the y-axis, periodic boundary conditions along the x-axis). Here $\Theta$ is the phase between neighboring site-rings along the axis with periodic boundary conditions. The edge band is highlighted in grey. (b) A schematic of the unit cell of the lattice. }
  \label{fig:bandstructure}
\end{figure}

Figure \ref{fig:bandstructure} shows the calculated band structure of a semi-infinite (periodic in one axis, finite in the other) AQH lattice. A $2J$-wide edge band region (highlighted in grey) resides between two bulk bands. Unlike the bulk modes, which lack well-defined momentum, there are two unidirectional edge bands with opposite pseudospins. These bands travel in opposite directions and are robust against local disorder. \cite{leykam2018reconfigurable,mittal2019photonic}. Additionally, the lattice unit cell is depicted schematically.

\subsection{Estimation of Device Parameters}

In our second quantized formalism, the strength of interaction $\beta$ is related to the effective Kerr nonlinear strength $\gamma$ common in the literature as $\gamma=\frac{\beta}{\hbar \omega} \frac{n_0^2 L_{\rm eff}}{c^2}$, where $n_0$ is the refractive index, $L_{\rm eff}$ is the circumference of the individual ring, and $\omega_0$ is the pump frequency. In order to estimate $\gamma$, we also require the nonlinear index $n_2$ and the effective mode area $A_{\rm eff}$. Here, a value of $n_2 = 2.4 \times 10^{-19}$~m$^2$ W$^{-1}$ is used \cite{Ikeda:08}. The refractive index of SiN at a wavelength of 1550 nm is taken as $n_0=2.00$ \cite{LukeSiN}. A value of $A_{\rm eff}= 0.9 \times 10^{-12}$~m$^2$ was calculated using FDTD simulation and $L_{\rm eff}=174~\mu \rm{m}$. With these values, we calculate $\gamma$ as:

\begin{equation}
\gamma = \frac{\omega_0 n_2}{c A_{\rm eff}} = 1.08~ \rm{W^{-1}m^{-1}}  \label{eqn:gamma}
\end{equation}

\begin{figure}
  \centering
  \includegraphics[width=0.9\textwidth]{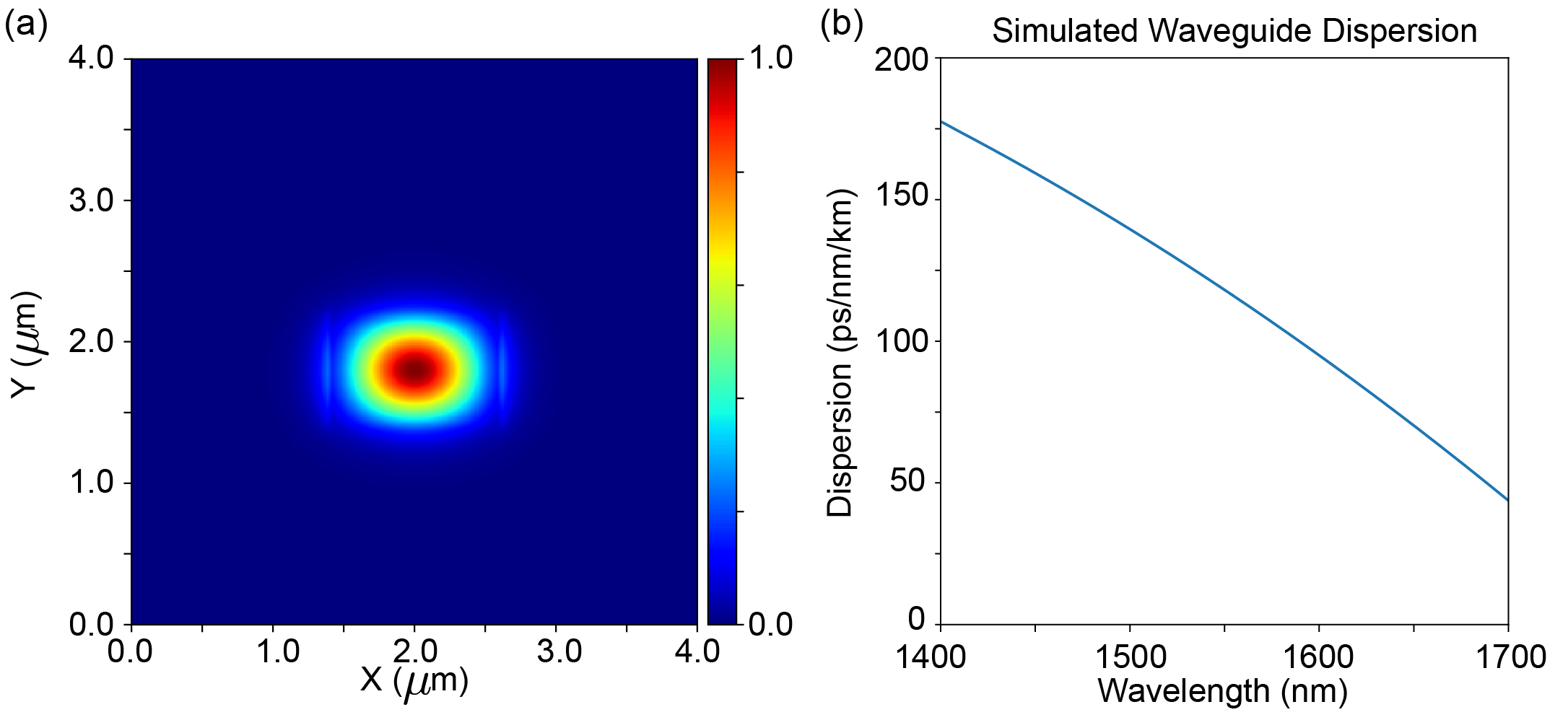}
  \caption{\textbf{Modal cross-section.}(a) Simulated electric field intensity in normalized units for a SiN waveguide cross-section of 1200 nm wide by 800 nm thick. Embedded in \ce{SiO2} cladding. (b) Simulated dispersion profile for the waveguide.}
  \label{fig:mode}
\end{figure}

We note that the definition of effective Kerr nonlinear strength $\gamma$ in this paper is common in the literature but differs from what was used in reference \cite{mittal2021topological}, in which it was taken as ${\omega_0 c n_2}/{n_0^2 V_{\rm eff}}$. The simulated mode profile and dispersion are displayed in Figure~\ref{fig:mode}, using the dependence of the refractive index on wavelength reported in \cite{LukeSiN}.

The relevant parameters used in theoretical modeling are estimated from our device as follows. $J$, the coupling strength between rings is estimated by comparing the measured drop spectrum of the lattice with simulation in the linear regime. The full bandwidth of edge and bulk bands is estimated to be approximately 8$J$ in simulation, while the edge band alone is approximately 2$J$. Using this, we arrive at an estimate of $J \approx2\pi\times$25 GHz for our device.  \hl{In order to approximate the splitting of individual edge modes,  we divide the bandwidth (2$J$) by the number of individual edge modes given the size of the lattice (20). This yields a splitting of $2\pi\times$2.5 GHz, which is 20 pm.}

The extrinsic coupling rate $\kappa_{\rm ex}$ and the intrinsic decay rate $\kappa_{\rm in}$ are estimated based on single-rings coupled to two waveguides, also known as add-drop filters (ADFs). Using the single-mode approximation, the transmission spectrum of the through port of an ADF with a given $\kappa_{\rm ex}$ and $\kappa_{\rm in}$ is a Lorentzian function:

\begin{equation}
     T=\frac{(\omega-\omega_0)^2+(\kappa_{\rm ex}^{\rm O}-\kappa_{\rm ex}^{\rm I}+\kappa_{\rm in})^2}{(\omega-\omega_0)^2+(\kappa_{\rm ex}^{\rm O}+\kappa_{\rm ex}^{\rm I}+\kappa_{\rm in})^2}.
\end{equation}

\begin{figure}
  \centering
  \includegraphics[width=.4\textwidth]{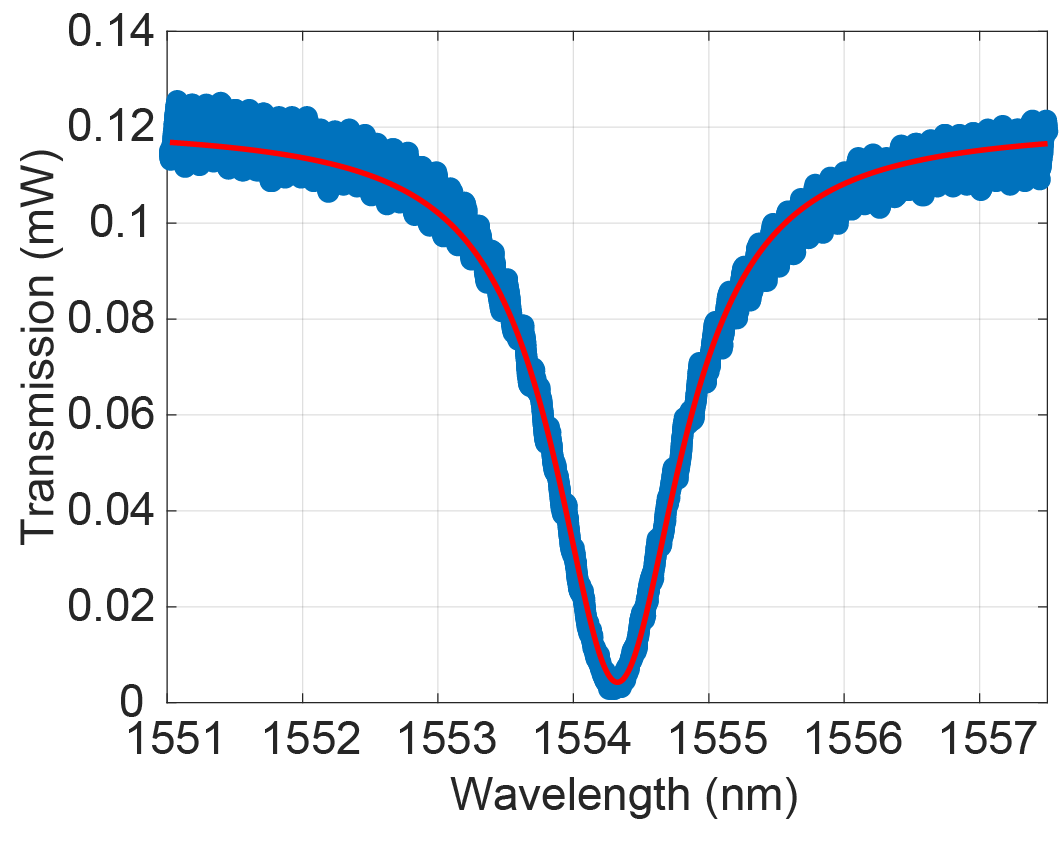}
  \caption{\textbf{Through spectrum of an ADF around one resonance.} Blue: Experimental data. Red: Lorentzian fitting.}
  \label{fig:adf}
\end{figure}

Here $\kappa_{\rm ex}^{\rm I}$ ($\kappa_{\rm ex}^{\rm O}$) is the extrinsic coupling to the input (output) waveguide. In the case of our lattice, due to identical coupling gaps, $\kappa_{\rm ex}^{\rm I}=\kappa_{\rm ex}^{\rm O}=\kappa_{\rm ex}$. By measuring the transmission from the through port (see Figure~\ref{fig:adf}) and fitting the curve to the above Lorentzian we extract the values $\kappa_{\rm ex} \approx2\pi\times$30 GHz and $ \kappa_{\rm in} \approx2\pi\times$2 GHz. This corresponds to a single-ring loaded quality factor of about 1500\hl{, and an intrinsic quality factor of about 50,000}.

\hl{We note that the relatively low quality factor is a result of the large value of $J$. This value is chosen in particular due to the role of disorder in the fabrication, which results in the detuning of the resonance frequencies of rings with identical parameters. Specifically, our topological devices are robust as long as the disorder in ring resonance frequencies is small compared to the width of the topological bandgap (2J). Additionally, topological lattices of this nature have been shown to be more robust to fabrication disorder than other systems, such as 1D chains of rings }\cite{mittal2014topologically}. \hl{While frequency comb generation has been studied theoretically in 1D chains of rings }\cite{Tusnin2023}\hl{, such structures are expected to be highly susceptible to fabrication disorder.} 

\hl{A propagation loss of -6.2 dB through the topological edge band in the CW direction through the lattice is estimated by comparing the through port off-resonant transmission with the on-resonant drop port transmission in the linear regime. }

\subsection{Measurement Setup and Methods}

For measurements in the linear regime, we couple a continuous-wave tunable laser to the input port (via edge couplers) and sweep the wavelength. Concurrently, the output power is measured at the drop port across the lattice with a power meter. For the measurement of the group delay of the drop port transmission, we use an optical vector network analyzer. In both cases, polarization is controlled with a standard 3-paddle polarization controller at the laser output.

\begin{figure}[h]
  \centering
  \includegraphics[width=0.90\textwidth]{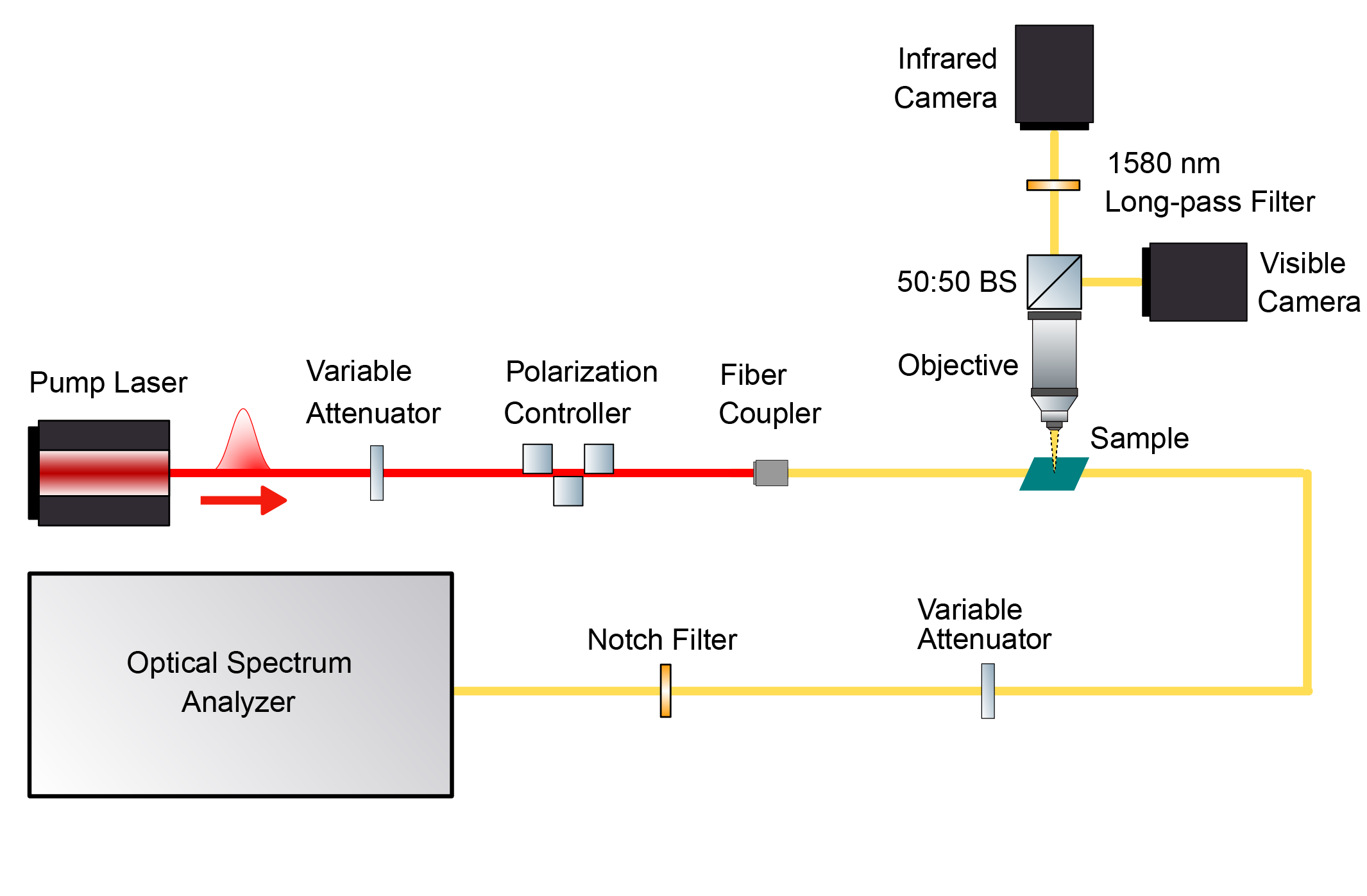}
  \caption{\textbf{Detailed schematic of the nonlinear measurement setup.} A tunable telecom pulsed laser is sent through a variable attenuator and polarization controller before being fiber-coupled and sent into the SiN device. The output of the device is then fiber-coupled and optionally attenuated by a second variable attenuator and notch filter for pump removal before being sent to the Optical Spectrum Analyzer. The chip is also imaged from above with a 10x objective, followed by a 50:50 beamsplitter. One optical path is sent to a visible camera, while the other is filtered by a 1580 nm long-pass filter and sent to an IR-sensitive camera.}
  \label{fig:setup}
\end{figure}
A schematic of the experimental setup for nonlinear measurement is shown in Figure~\ref{fig:setup}. For these measurements, we couple a pulsed tunable laser to a free-space optical setup including a variable attenuator and a polarization controller comprised of a quarter, half, and quarter wave-plate. The output is then coupled into a short tapered fiber and edge coupled into the SiN chip via the input port. \hl{Coupling losses are estimated to be 2-3 dB per coupler.} The pump and comb output are collected from the drop port with another tapered fiber and optionally attenuated or filtered prior to being sent into an optical spectrum analyzer.

For imaging, we collect out-of-plane scattering from the SiN chip with a 10x objective lens \hl{with a numerical aperture of 0.28.} The image is sent through a 50:50 beamsplitter to a visible wavelength camera as well as an infrared (IR) sensitive camera. A 1580 nm long-pass filter is included prior to the IR camera to filter out the pump laser.

\subsection{Pump Laser Characterization}
The pump laser used in the nonlinear measurements described in this work is a  tunable pulsed laser. The pulse duration is designed to be approximately 5 ns with a repetition rate of 250 kHz. This relatively long pulse duration (compared to the longest timescales of the lattice dynamics) and low repetition rate provide access to a high power quasi-continuous wave regime. In particular, operation in this regime allows power requirements to be satisfied and detrimental thermal effects to be minimized while numerical modeling using a single frequency pump remains applicable.

\begin{figure}[h]
  \centering
  \includegraphics[width=0.99\textwidth]{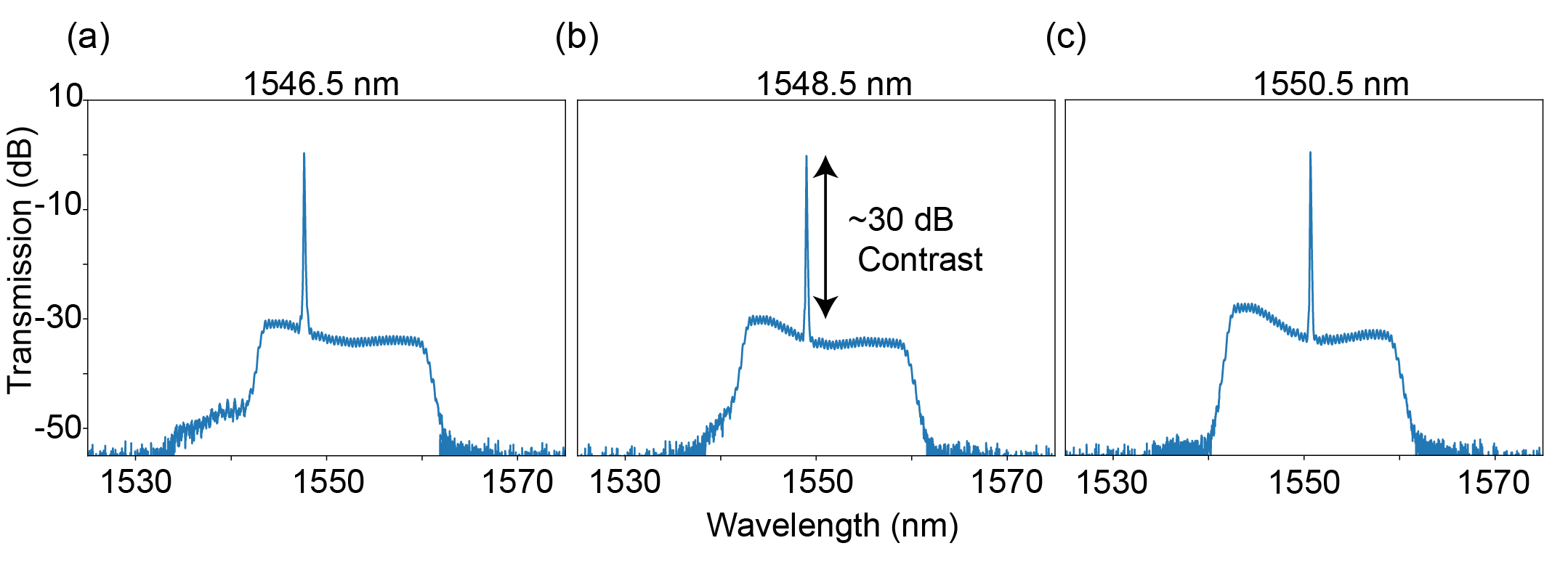}
  \caption{\textbf{Laser characterization.} The unfiltered spectrum of the pump laser at approximately (a) 1546.5 nm, (b) 1548.5 nm, (c) and 1550.5 nm. }
  \label{fig:Laser}
\end{figure}

Three pump laser spectra are shown in Figure~\ref{fig:Laser} for three different central wavelengths. As can be seen, the laser background remains approximately 30 dB suppressed from the laser peak regardless of the pump central wavelength. \hl{The spectral linewidth of the pump is approximately 3 pm.}

\begin{figure}[h]
  \centering
  \includegraphics[width=0.4\textwidth]{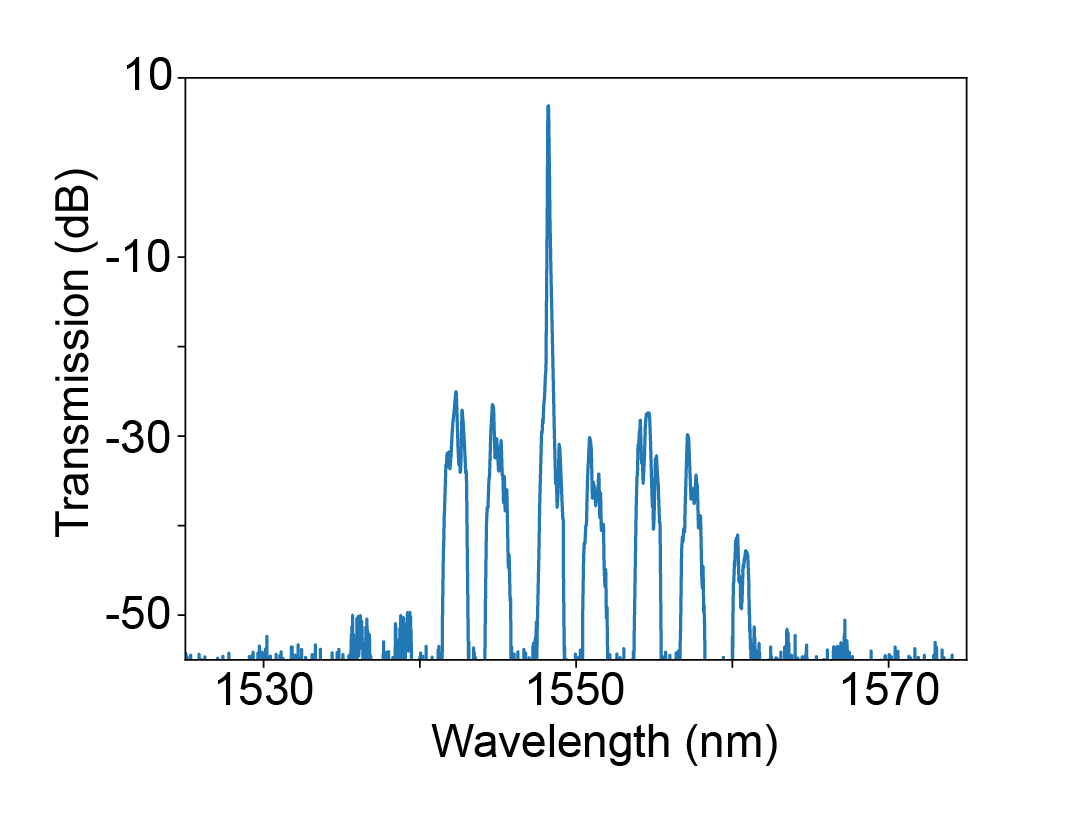}
  \caption{\textbf{Laser background drop spectrum.} The output drop spectrum of an AQH lattice when the pump is detuned and below the comb formation threshold. }
  \label{fig:Laser2}
\end{figure}

The same laser envelope can be observed when sending the pump through a topological lattice and collecting the drop port spectrum. The output spectrum of an identical AQH lattice as the one used in the nonlinear measurements in this work with a pump power below the comb formation threshold is shown in Figure~\ref{fig:Laser2}. The broad laser background is transmitted through several edge bands as well as link-ring bands, producing a characteristic structure that can be observed in the main text. 

\subsection{Comb Contrast}
\begin{figure}[h]
  \centering
  \includegraphics[width=0.7\textwidth]{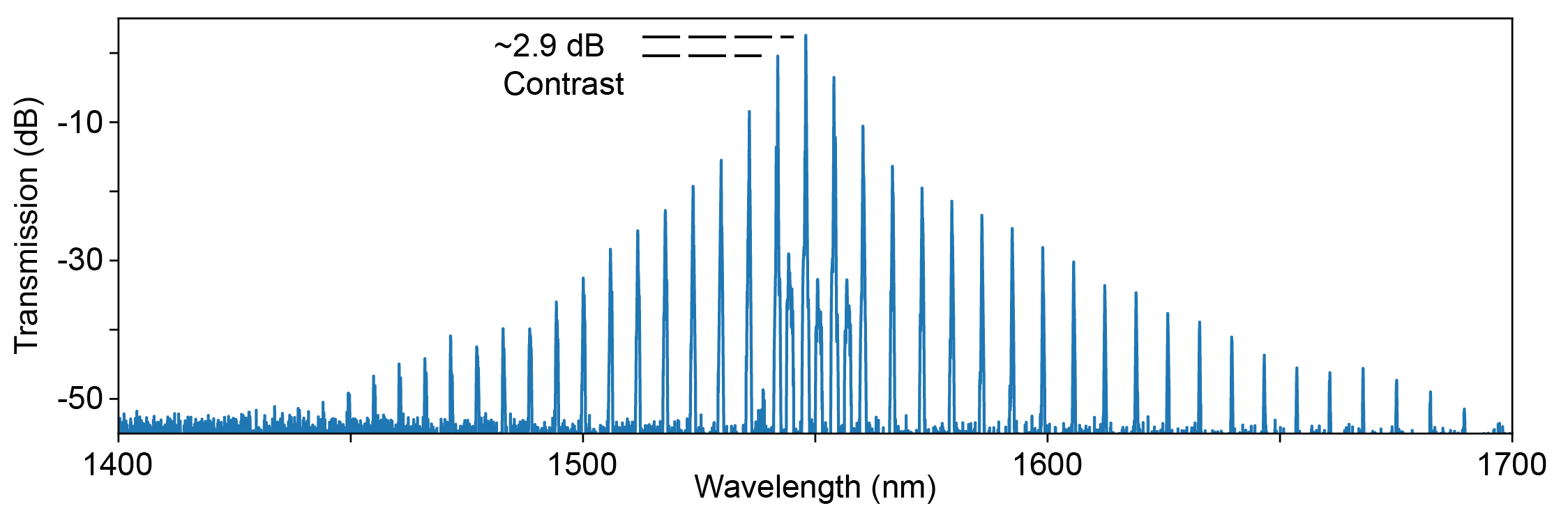} 
  \caption{\textbf{Comb Contrast. }\hl{Spectrum of a topological frequency comb comparable to that of Figure} \ref{fig:nonlinear}\hl{e with the pump laser included.}}
  \label{fig:contrast}
\end{figure}

Figure \ref{fig:contrast} \hl{shows a reference spectrum comparable to that of Figure} \ref{fig:nonlinear}\hl{e with the pump laser included. }

\subsection{Nesting Comparison}
\begin{figure}[h]
  \centering
  \includegraphics[width=0.65\textwidth]{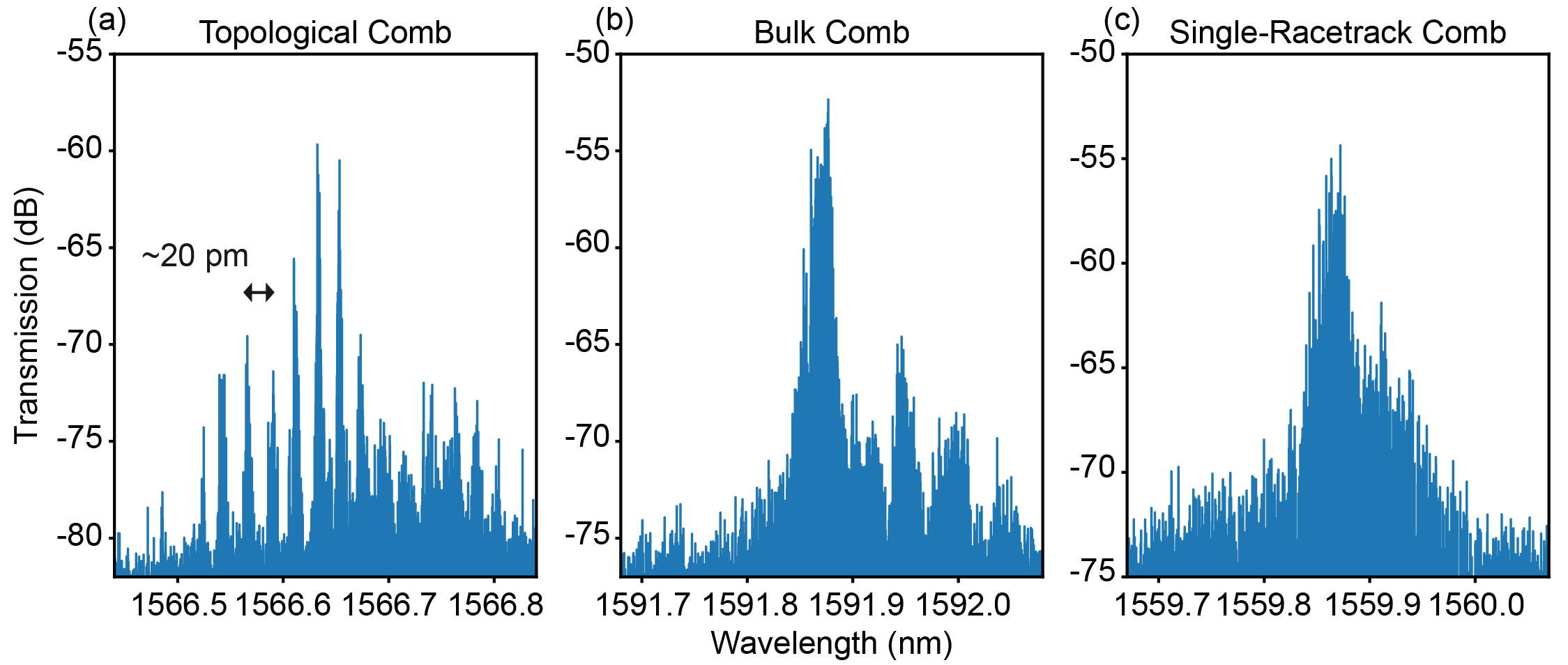} 
  \caption{\textbf{High-Resolution Spectra of Individual Comb Teeth.}\hl{ (a) A high-resolution comb tooth from the topological frequency comb, displaying nesting. (b) A high-resolution comb tooth from a bulk comb. (c) A high-resolution comb tooth from a single-racetrack comb.}}
  \label{fig:linewidth}
\end{figure}

\hl{In order to contrast the high-resolution spectra in Figure}~\ref{fig:nesting} \hl{with more typical, non-nested frequency combs, we generate a bulk comb as described in the main text, as well as a frequency comb in a single-racetrack resonator with identical dimensions as the site-ring resonators in the topological lattice, shown in Figure}~\ref{fig:linewidth} \hl{. The on-chip peak power used for the bulk comb is approximately 122 W with a pump wavelength of 1547.43 nm. The single-racetrack resonator is coupled in an add-drop filter configuration, where there are two bus waveguides. Notably, the coupling gaps here are increased from 300 nm to 600 nm in order to increase the loaded Q factor from $\approx1500$ to $\approx21,000$. For the single-racetrack comb, the pump wavelength is 1547.32 nm and the peak power is approximately 104 W. Linewidths for the main peaks of each comb tooth are approximately 3 pm, 13 pm, and 18 pm respectively. }

\newpage

\newpage
\subsection{Generation and Spectra of Imaged Modes}

\begin{figure}[h]
  \centering
  \includegraphics[width=0.6\textwidth]{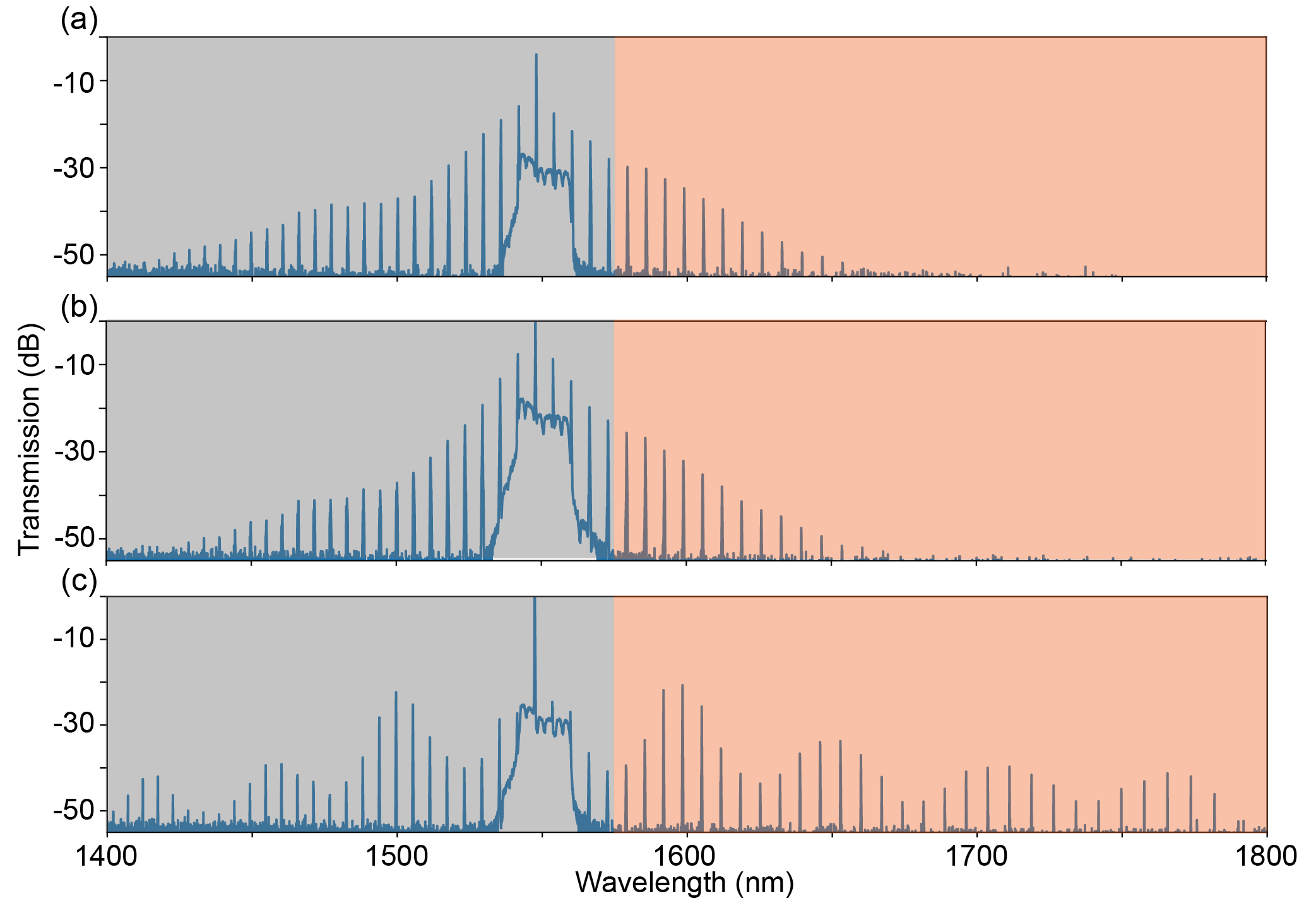}
  \caption{\textbf{CCW, CW, and bulk comb spectra.} The generated comb spectrum when the pump is tuned in resonance with a (a) CCW mode,  (b) CW mode, and (c) bulk mode. Filtered wavelengths are highlighted in grey, while imaged wavelengths are highlighted in red.}
  \label{fig:Img_spectra}
\end{figure}

For the CCW and CW topological frequency comb images in Figure~\ref{fig:imaging}, the peak pump power in the waveguide was approximately 92 W and 100 W respectively. The wavelength was tuned through the edge band to a value of 1547.97 nm. In order to generate a bulk comb, the peak pump power in the waveguide was approximately 125 W and the laser was tuned onto resonance with a bulk mode at 1547.43 nm. Through port spectra of each of these combs are displayed in Figure~\ref{fig:Img_spectra}. We note that while the pump power used here for the bulk mode is significantly higher than that of the topological combs, a broader systematic study would be required to determine the minimum power required to generate frequency combs in bulk modes of the lattice. Due to the highly variable nature of these bulk modes, such a study is outside the scope of the present work. 

\newpage
\end{document}